\documentclass[graybox,natbib,nosecnum]{svmult}
\bibpunct{(}{)}{;}{a}{}{,} 

\pdfoutput=1   
\usepackage{amsmath}
\usepackage{amssymb}
\usepackage{mathptmx}       
\usepackage{helvet}         
\usepackage{courier}        
\usepackage{type1cm}        

\usepackage{makeidx}         
\usepackage{graphicx}        
\usepackage{multicol}        
\usepackage[bottom]{footmisc}
\usepackage[normalem]{ulem}	
\usepackage{hyperref}  
\usepackage{subcaption}

\usepackage{soul}   

\newcommand*\apj{ApJ}

\makeindex             

\begin{document}

\title*{Populations of planets in multiple star systems}
\author{David V. Martin}
\institute{David V. Martin \at Fellow of the Swiss National Science Foundation at the University of Chicago \\  \email{davidmartin@uchicago.edu}}
%
%
\maketitle

\abstract{Astronomers have discovered that both planets and binaries are abundant throughout the Galaxy. In combination, we know of over 100 planets in binary and higher-order multi-star systems, in both circumbinary and circumstellar configurations. In this chapter we review these findings and some of their implications for the formation of both stars and planets. Most of the planets found have been circumstellar, where there is seemingly a ruinous influence of the second star if  sufficiently close ($\lesssim50$ AU).  Hosts of hot Jupiters have been a particularly popular target for binary star studies,  showing an enhanced rate of stellar multiplicity for moderately wide binaries (beyond $\sim 100$ AU).  This was thought to be a sign of Kozai-Lidov migration, however recent studies have shown this  mechanism to be too inefficient to account for the majority of hot Jupiters. A couple of dozen circumbinary planets have been proposed around both main sequence and evolved binaries. Around main sequence binaries  there are preliminary indications that the frequency of gas giants is as high as those around single stars. There is however a conspicuous absence of circumbinary planets around the tightest main sequence binaries with periods of just a few days, suggesting a unique, more disruptive formation history of such close stellar pairs.}




\section{Introduction }

 It is known that roughly half  of Sun-like stars exist  in multiples and about a third in binaries \citep{heintz69,duquennoy91,raghavan10,tokovinin14}. It is also known that extra-solar planets are highly abundant, with most stars  hosting at least one planet \citep{howard10,mayor11,petigura13}. The next step is to connect the two concepts and pose the question of planets in binaries. Such planets are often thought of as exotic examples of nature's diversity. However, considering the ubiquity of both planets and binaries throughout the Galaxy, the question of their coupled existence is in fact natural. 

We first cover a few important aspects of stellar multiplicity and the configurations, stability and dynamics of planets in binaries. The rest of this chapter is devoted to analysing the observed populations of planets in binaries. Some of the implications for planet formation are also discussed.


\subsection{Stellar multiplicity}\label{subsec:stellar_multiplicity}

\begin{figure}  
\captionsetup[subfigure]{labelformat=empty}
\begin{center}  
	\begin{subfigure}[b]{0.49\textwidth}
		\includegraphics[width=\textwidth]{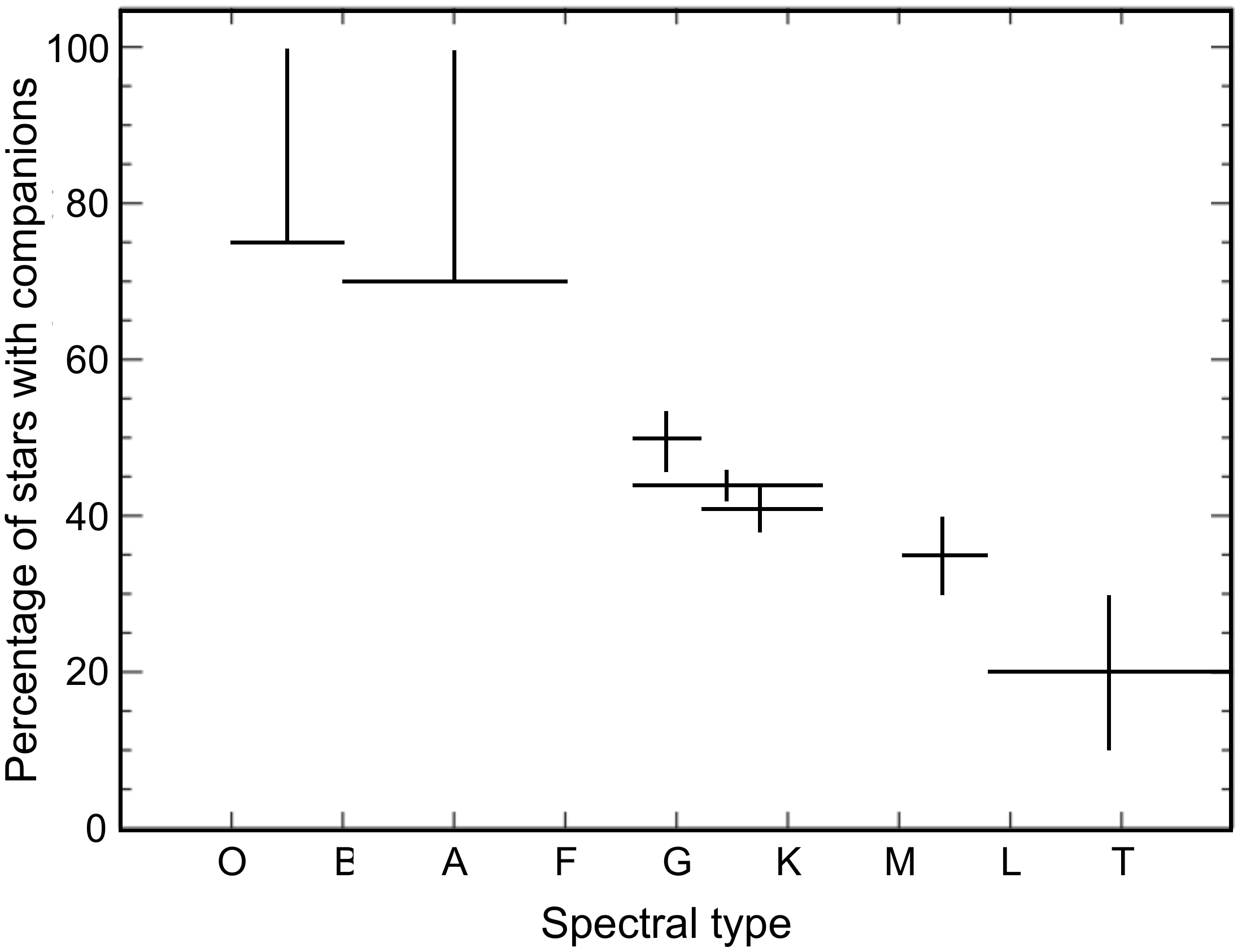}  
	\end{subfigure}
	\begin{subfigure}[b]{0.49\textwidth}
		\includegraphics[width=\textwidth]{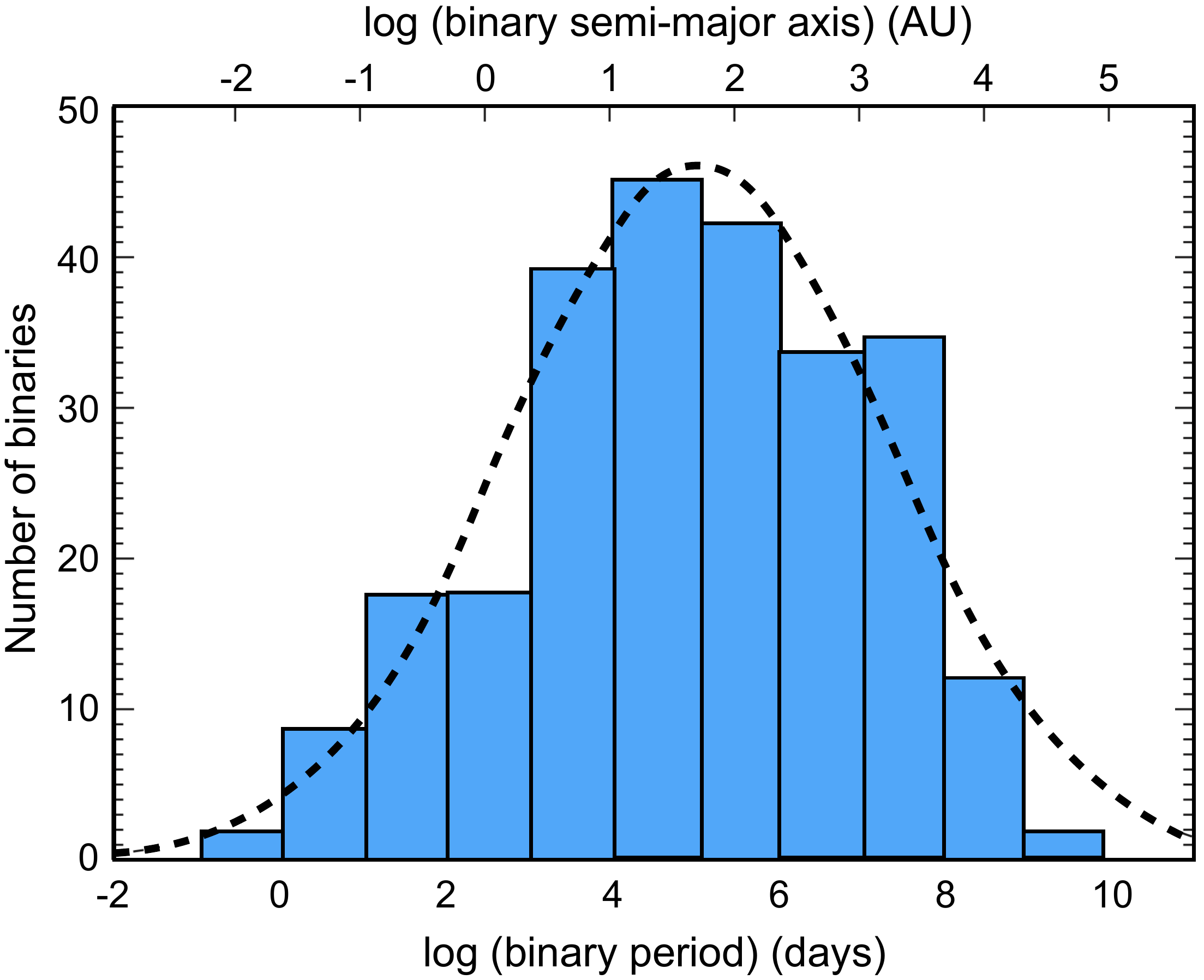}  
	\end{subfigure}	
	\caption{Left: stellar companion percentage as a function of spectral type. Right: period distribution of observed binaries, with a semi-major axis distribution calculated assuming a mass sum of $1.5M_{\odot}$, which is the average observed value. The dashed line is a fitted log-normal distribution with a mean of $\log T_{\rm bin}=5.03$ and a standard deviation of $\sigma_{\log T_{\rm bin}}=2.28$. Both figures are adapted from \citet{raghavan10}, with the data taken from sources listed in that paper.}
	\label{fig:stellar_multiplicity}  
\end{center}  
\end{figure}

The seminal work of \citet{raghavan10} draws upon binary  and higher-order multi-star systems discovered with a variety of techniques. Two of the most important results are the multiplicity rate of stars and the separation distribution for binaries. These two results are shown in Fig.~\ref{fig:stellar_multiplicity}. For FGK stars that are typically considered for exoplanet surveys $\sim$40-50\% of stars have additional companions. The multiplicity is higher for more massive stars and lower for less massive. For binary stars the distribution of separations can be reasonably fitted by a log-normal  function with a mean of 293 years. In terms of semi-major axis, this corresponds to roughly 50 AU for a mass sum $M_{\rm A}+M_{\rm B}=1.5M_{\odot}$.  This distribution of separations is calculated using primaries of all masses. When split into different primary spectral types, the semi-major axis distribution grows wider as a function of increasing primary mass.

\subsection{Orbital configurations}\label{subsec:configurations}

\begin{figure}
\includegraphics[width=0.9\textwidth]{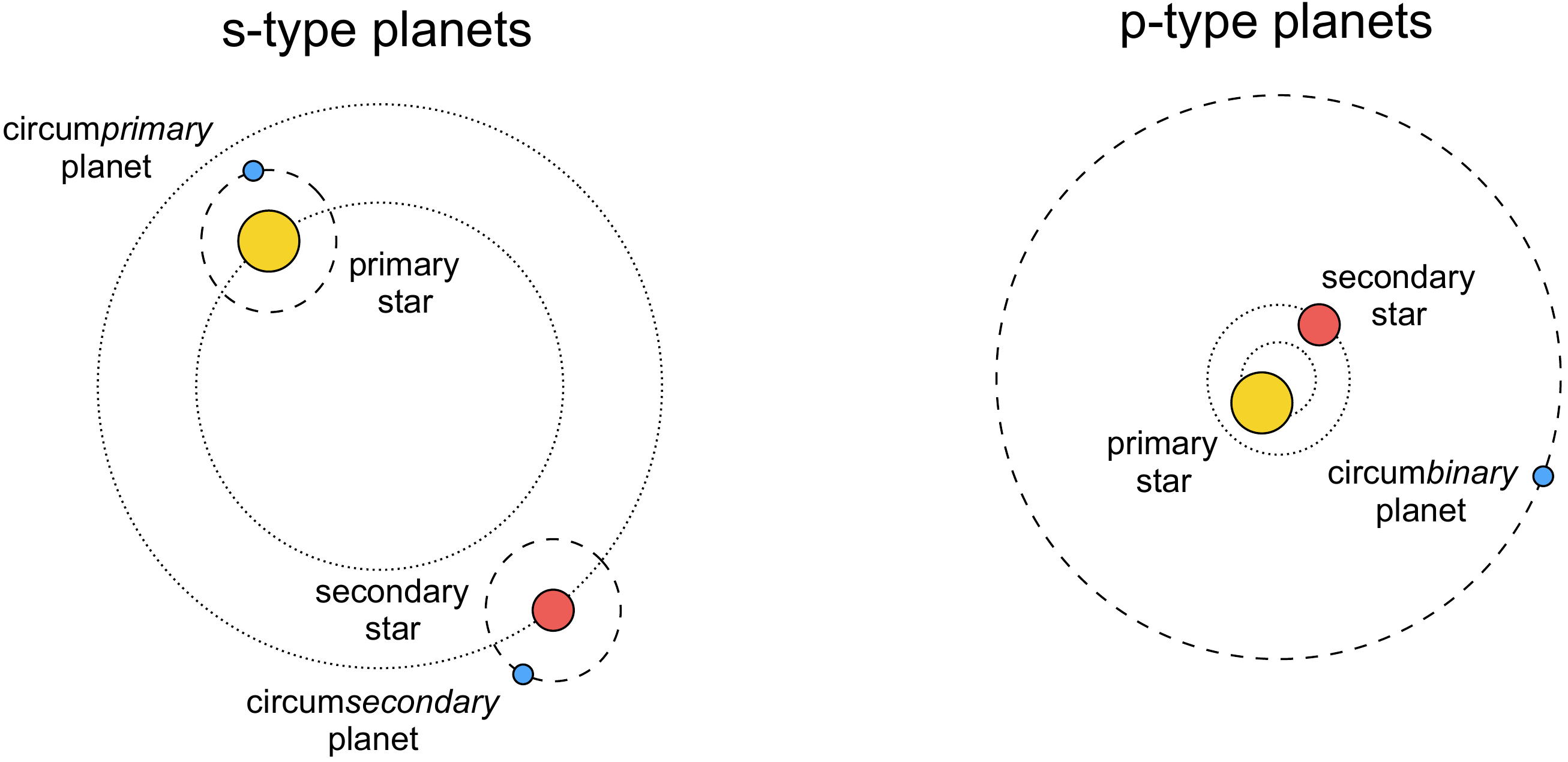}
\caption{Left: circumstellar ``s-type'' planets in binaries, individually around both primary and secondary stars. Right: circumbinary ``p-type'' orbits in binaries collectively around both stars.}
\label{fig:binary_orbits}
\end{figure}

 There are two types of orbits in which planets have been discovered in binary star systems. First, the planet may have a wider orbit than the binary ($a_{\rm p}>a_{\rm bin}$) and orbit around the barycentre of the inner binary. This is known as a circumbinary or ``p-type'' planet. Alternatively, the planet may have a smaller orbit than the binary ($a_{\rm p} < a_{\rm bin}$) and only orbit around one component. This  is known generally as a circumstellar or ``s-type'' planet, or as a circumprimary or circumsecondary planet as a function of which star is being orbited\footnote{These terms were first coined in \citet{dvorak86} and stand for ``planet-type'' and ``satellite-type''.}. These configurations are illustrated in Fig.~\ref{fig:binary_orbits}.  Other, more exotic orbits in binaries have been considered, such as trojan planets near L4 and L5 \citep{dvorak86,schwarz15} and halo orbits near L1, L2 and L3 \citep{howell83}. No such planets have been discovered though.

\subsection{Orbital stability}\label{subsec:stability}

There is a limit to where a planet may have a stable orbit in a binary star system. This has a profound effect on the observed populations, by carving away unstable regions of the parameter space. Much of the work to derive three-body stability limits was undertaken even before planets were discovered in binaries \citep{ziglin75,black82,dvorak86,eggleton95,holman99,mardling01,lohinger03,mudryk06,doolin11}. The classic method has been to run numerical $N$-body simulations over a parameter space and determine regular and chaotic domains. The often-quoted work of \citet{holman99} used this method to derive empirical stability limits for both circumbinary and circumstellar planets. 

 Circumbinary planets have stable orbits beyond $a_{\rm crit}$,

\begin{align}
\begin{split}
\label{eq:circumbinary_stability_HW}
\frac{a_{\rm crit}}{a_{\rm bin}} &= 1.60 + 5.10e_{\rm bin}  - 2.22e_{\rm bin}^2 + 4.12\mu_{\rm bin} - 4.27 e_{\rm bin} - 5.09 \mu_{\rm bin}^2 + 4.61e_{\rm bin}^2\mu_{\rm bin}^2,
\end{split}
\end{align}
 where $a_{\rm bin}$ is the semi-major axis of the binary, $e_{\rm bin}$ is the eccentricity of the binary and $\mu_{\rm bin} = M_{\rm B}/(M_{\rm A}+M_{\rm B})$ is the reduced mass of the binary. This does not account for eccentric or misaligned planets or resonances, which can create  islands of both stability and instability \citep{doolin11}. For circumstellar orbits the widest planet orbit $a_{\rm crit}$ is

\begin{align}
\begin{split}
\label{eq:circumstellar_stability_HW}
\frac{a_{\rm crit}}{a_{\rm bin}} &= 0.464 - 3.80\mu_{\rm bin} -0.631e_{\rm bin}  + 0.586 e_{\rm bin}\mu_{\rm bin} + 0.150 e_{\rm bin}^2 -0.198\mu_{\rm bin}e_{\rm bin}^2.
\end{split}
\end{align}

For details, including error bars on the coefficients of Eqs.~\ref{eq:circumbinary_stability_HW} and ~\ref{eq:circumstellar_stability_HW} see \citet{holman99}.

\subsection{Kozai-Lidov cycles}\label{subsec:kozai_lidov}

\begin{figure}  
\captionsetup[subfigure]{labelformat=empty}
\begin{center}  
	\begin{subfigure}[b]{0.49\textwidth}
		\includegraphics[width=\textwidth]{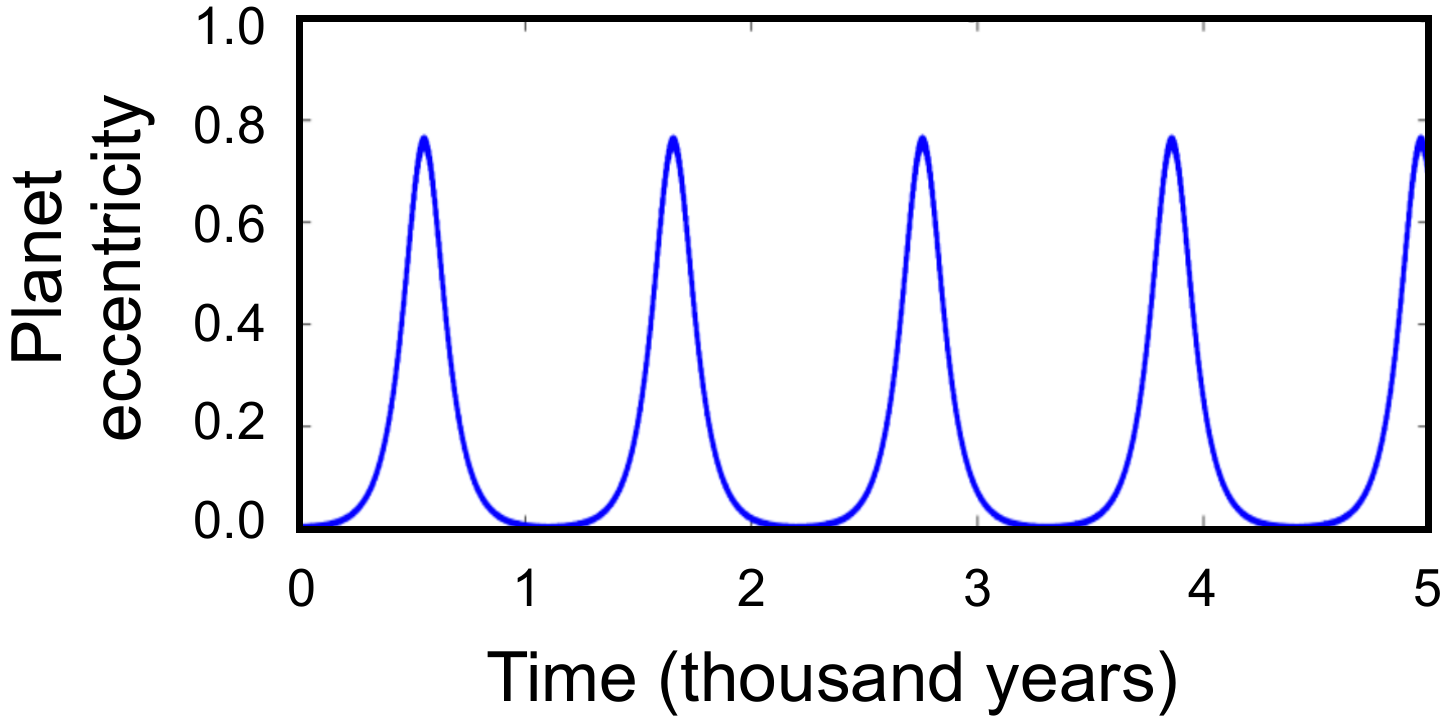}  
	\end{subfigure}
	\begin{subfigure}[b]{0.48\textwidth}
		\includegraphics[width=\textwidth]{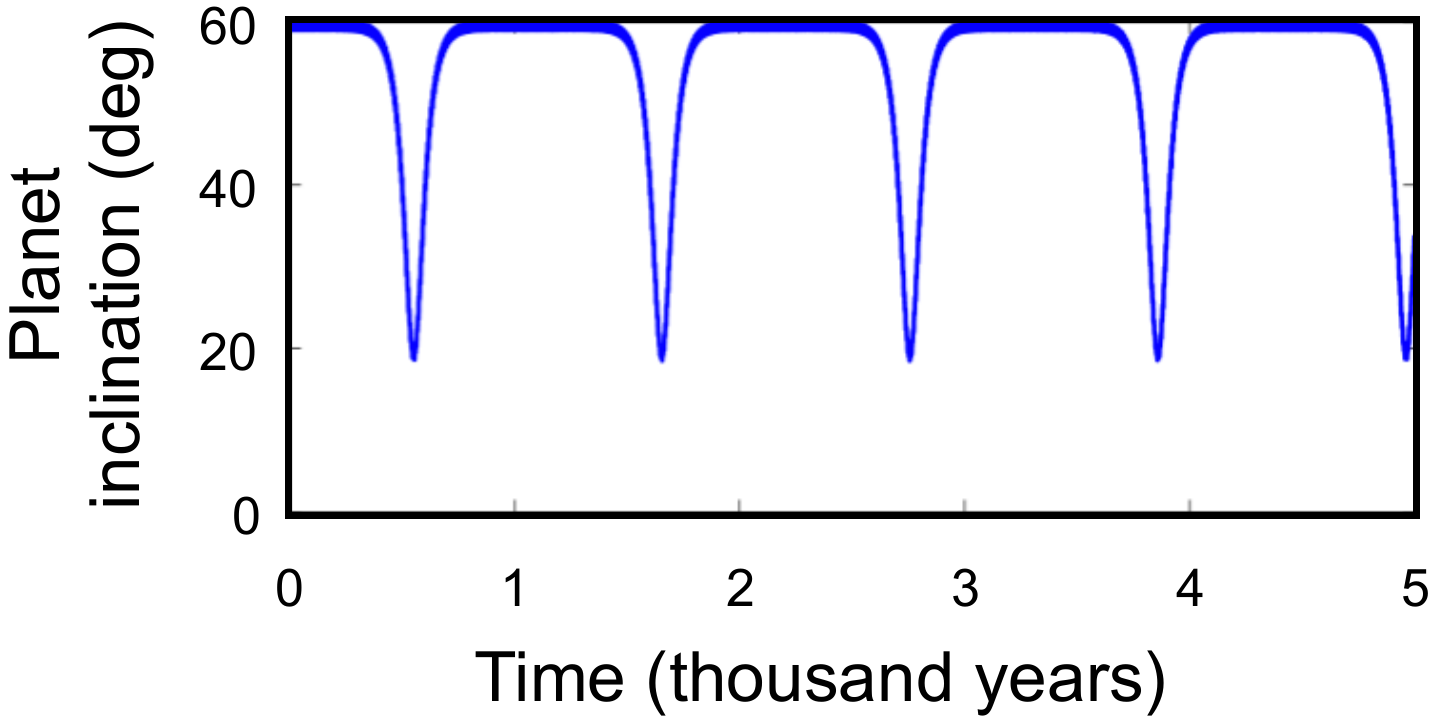}  
	\end{subfigure}	
	\caption{Example of Kozai-Lidov cycles for a  0.5 AU circumprimary planet in a 5 AU binary, showing the variation of $e_{\rm p}$ (left) and $I_{\rm p}$ (right). The planet is initially inclined by $60^{\circ}$ with respect to the binary's orbital plane.}
	\label{fig:kozai_lidov_example}  
\end{center}  
\end{figure}

For circumstellar planets in binaries one must consider the Kozai-Lidov effect, which is named after the pioneering work of \citet{lidov61,lidov62,kozai62}. If the planet and binary orbits are misaligned between $39^{\circ}$ and $141^{\circ}$ then there is a secular oscillation of both the planet's eccentricity, $e_{\rm p}$, and its inclination with respect to the binary, $I_{\rm p}$. An example is shown in Fig.~\ref{fig:kozai_lidov_example}. An initially circular circumstellar planet obtains a maximum eccentricity of 

\begin{equation}
e_{\rm p,max} = \sqrt{1-\frac{5}{3}\cos^2I_{\rm p,0}},
\end{equation}
where $I_{\rm p,0}$ corresponds to the planet's inclination at $e_{\rm p}=0$. This is  derived to quadrupole order, under the assumption that the outer orbit (here the binary) carries the vast majority of the angular momentum. The outer eccentricity and inclination do not change.  More general equations that can be applied to any inner and outer angular momenta were derived in \citet{lidov76,naoz13,liu15}. For {\it circumbinary} planets, where the outer angular momentum is typically negligible, the Kozai-Lidov effect practically disappears \citep{migaszewski11,martin16}.

%

\section{Discoveries and analysis}\label{sec:observations}

\begin{figure}
\includegraphics[width=0.99\textwidth]{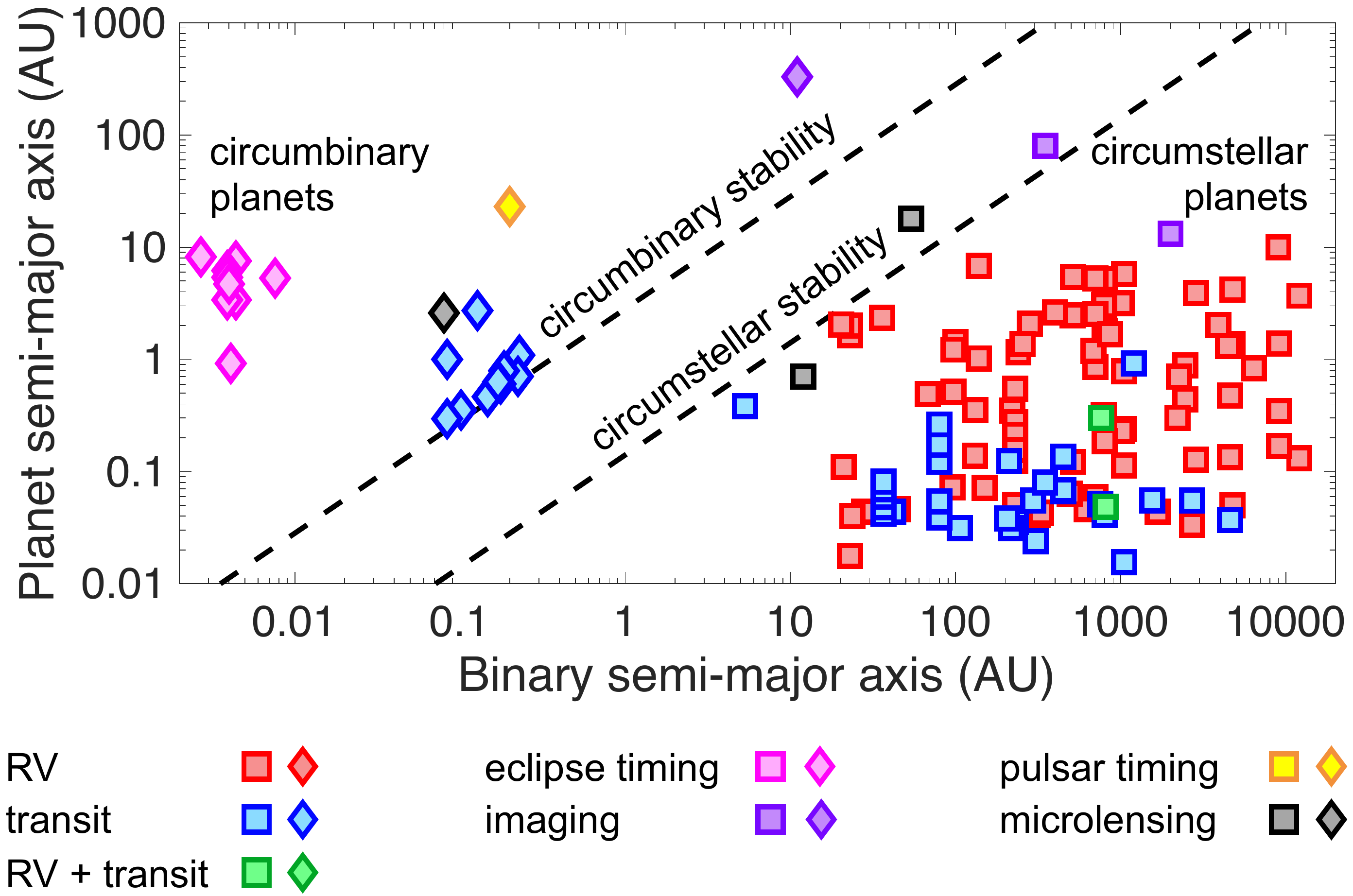}
\caption{Planets in multi-star systems. Circumbinary planets are denoted by diamonds and circumstellar planets by squares. The different colours indicate the discovery technique for the planet, not the binary. The circumbinary and circumstellar stability limits are calculated using Eqs.~\ref{eq:circumbinary_stability_HW} and ~\ref{eq:circumstellar_stability_HW}, respectively, with $M_{\rm A}=1M_{\odot}$, $M_{\rm B}=0.5M_{\odot}$. For circumbinaries $e_{\rm bin}=0.15$ (mean for transiting discoveries) and for circumstellar planets $e_{\rm bin}=0.5$ (representative of wider binaries, \citealt{tokovinin16}).}
\label{fig:all_planets}
\end{figure}

Despite thousands of exoplanet discoveries to date, only a small fraction are known to exist in multi-star systems. This may seem surprising given the  frequency of binary stars, but there have however been historical biases and strategies against finding planets in such systems \citep{eggenbergerudry07,wright12}. 

A catalog of planets in binaries and multi-star systems is maintained by Richard Schwarz (\citealt{schwarz16}, \url{http://www.univie.ac.at/adg/schwarz/multiple.html}). As of May 2017 it lists 113 planets in 80 binaries and an additional 33 planets in 24 triple and higher order stellar systems. A comparison between binary and higher-order stellar systems is beyond the scope of this chapter, although we note that the first planet found in a multi-star system was found in a triple (16 Cyg, \citealt{cochran97}).  The closest exoplanet known also exists in a triple (Proxima Cen, \citealt{angladaescude16}). We only know of two planets  which exist in a circumbinary configuration but also have outer stellar companions, and hence possess both p-type and s-type orbits (PH-1/Kepler-64, \citealt{schwamb13,kostov13} and HW Virginis, \citealt{lee09}).

In Fig.~\ref{fig:all_planets} is a plot of the planet and binary semi-major axes for all  systems in the Schwarz catalog with these values recorded. For planets in multi-star systems $a_{\rm bin}$ is the separation to the closest stellar companion to the host star. In triple and higher-order systems the closest stellar companion may itself be a binary. HW Virginis and PH-1/Kepler-64 are plotted as circumbinary systems.

This figure demonstrates that the circumbinary and circumstellar planets are naturally separated by the two stability limits, with roughly eight of each type near the respective stability boundary. According to the plot, two circumstellar planets are seemingly outside of the stable parameter space: OGLE-2008-BLG-092L (black square, \citealt{poleski14}) and HD 131399 (purple square, \citealt{wagner16}). However, in both cases the orbit may be stable for binary eccentricities less than the value of 0.5 used to demarcate the stability limit in Fig.~\ref{fig:all_planets}. Furthermore, \citet{nielsen17} present evidence that the planet in HD 131399 may in fact be a false-positive background star.

The circumstellar discoveries are more numerous than the circumbinaries so far, at a ratio of roughly 5:1. However since circumbinary discoveries are in their infancy this ratio is not meaningful. Because the two populations are seemingly distinct, we treat them in their own separate sections.


\subsection{Circumbinary planets}\label{subsec:circumbinary}


There have been many attempts with different techniques to find circumbinary planets. A general review of  circumbinary detection methods is provided in the chapter by Doyle \& Deeg. In Fig.~\ref{fig:all_planets} we see that two techniques have dominated the circumbinary landscape: transits and eclipse timing variations (ETVs). Welsh \& Orosz review the {\it Kepler} mission's search for transiting systems.  The chapter by Marsh covers  the proposed discoveries of planets around post-common envelope binaries uncovered by ETVs, although this technique is also applicable to main sequence binaries \citep{schwarz11}. The few remaining circumbinary discoveries have some from pulsar timing, microlensing and imaging. Three of the imaging circumbinary planets - SR 12 AB c, Ross 458 c and ROXs 42b are not displayed in Fig.~\ref{fig:all_planets} because they lack a value for $a_{\rm bin}$ in the Schwarz catalog (see \citealt{kraus14,bowler14} for more details on their characterisation).

The method of radial velocities (RVs), which has been highly productive for planets around single stars, is yet to yield a bonafide circumbinary planet. This is despite concerted efforts over the years (e.g. TATOOINE, \citealt{konacki05,konacki10}). A potential circumbinary planet in HD 202206 was proposed by \citet{correia05}, but later astrometry characterised it as a circumbinary brown dwarf \citep{fritz17}. Astrometry with {\it GAIA}  has the potential to find  massive new circumbinary planets at moderate separations (a few AU) and also confirm or deny some of the ETV  candidates \citep{sahlmann15}.

\begin{figure*}  
\begin{center}  
\includegraphics[width=\textwidth]{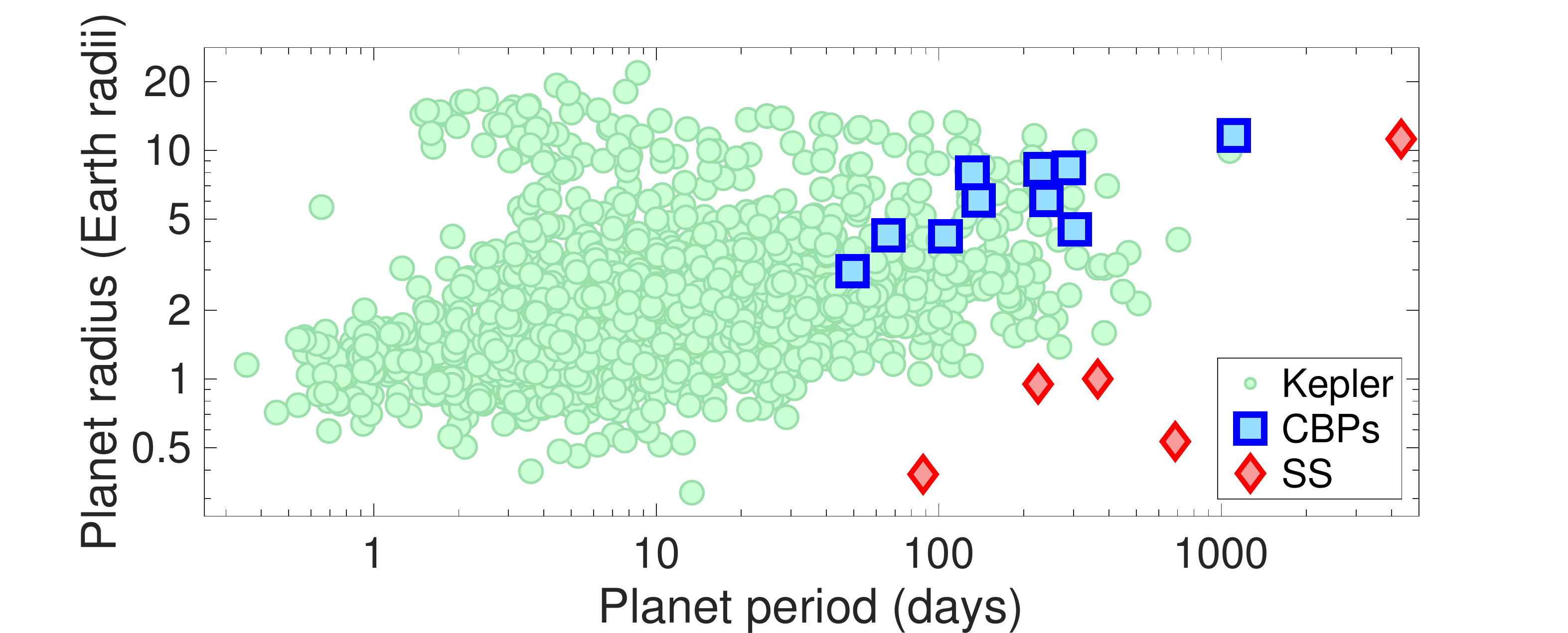}  
\caption{Period (in days) vs radius (in Earth radii) for the {\it Kepler} objects of interest around all stars (green circles), transiting circumbinary planets (CBPs, blue squares) and the five innermost Solar System planets (SS, red diamonds).}
\label{fig:circumbinary_radius_vs_period}
\end{center}  
\end{figure*}

\subsubsection{Observed trends}\label{subsubsec:circumbinary_trends}


When analysing the trends of circumbinary planets we largely stick to results of the {\it Kepler} transit survey. This is because it is the only sample that is both large enough for preliminary population studies and contains reliable discoveries, unlike the many caveats of the proposed ETV planets. Also by limiting ourselves to a single observing technique only a single observing bias needs to be accounted for.

 The smallest circumbinary planet discovered to date is $3R_{\oplus}$; the rest are all larger than Neptune. They also have periods between 49 and 1108 days, which span those in the inner Solar System and are considered long for transit surveys. This is evident in Fig.~\ref{fig:circumbinary_radius_vs_period} where the circumbinary planets populate the top right of the parameter space. Finding circumbinary planets at long periods is aided by a transit probability which, compared with that around single stars, is both higher and has a shallower dependence on orbital period \citep{schneider90,schneider94,martin15,li16,martin17}.


There are evidently two stark holes in the circumbinary population: small planets and short-period planets. The shallow depth of small planets lowers the detection efficiency, however Fig.~\ref{fig:circumbinary_radius_vs_period} demonstrates that discoveries of them around single stars have been plentiful. Furthermore, studies of single stars such as \citet{petigura13} have demonstrated that small super-Earth and Earth-sized exoplanets are much more frequent than larger planets. The discovery of small circumbinary planets must however overcome an additional challenge: a  unique transit timing signature. 

For planets around single stars one may phase-fold the data on a certain period to stack transits and build statistical significance. For circumbinary planets, the barycentric motion of the binary and variation of the planetary orbit result in transit timing variations on the order of $(T_{\rm p}T_{\rm bin}^2)^{1/3}/(2\pi)$ \citep{agol05,armstrong13}. This may be on the order of days, and hence significantly longer than the transit duration. This inhibits the effectiveness of phase-folding for circumbinaries. All of the discoveries to date were made by eye,  which is only effective when each individual transit is highly significant as in the case of giant planet transits.
\begin{figure*}  
\begin{center}  
\includegraphics[width=0.8\textwidth]{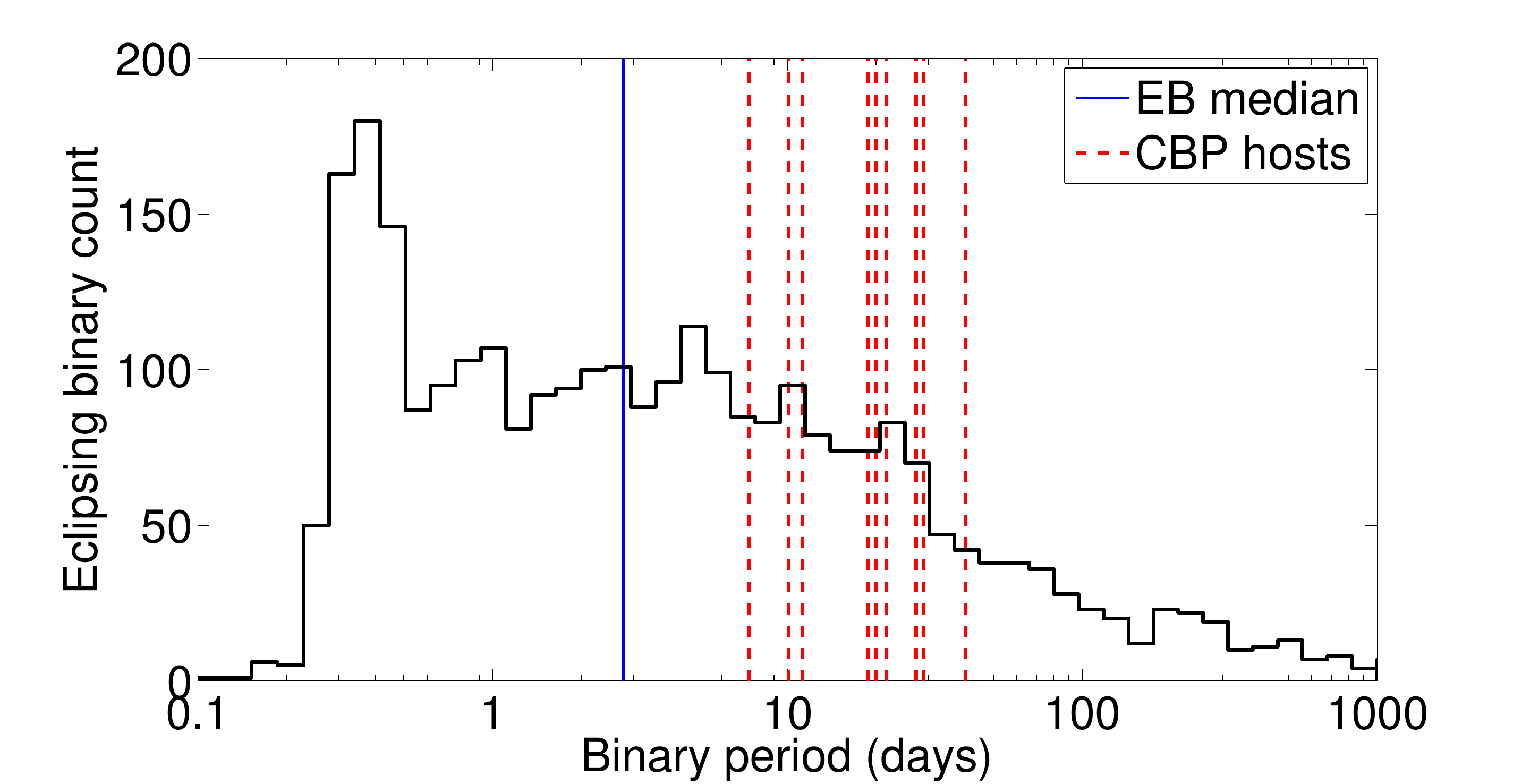}  
\caption{Histogram of the 2862 {\it Kepler} eclipsing binaries in black (\url{http://keplerebs.villanova.edu/} and first outlined in \citealt{prsa11}). The median at 2.8 days is denoted by a vertical blue solid line. The periods of binaries known to host transiting circumbinary planets are shown as vertical red dashed lines.}
\label{fig:circumbinary_EB_histogram}
\end{center}  
\end{figure*}

For the lack of $< 50$ day planets, there are two components. First, the stability limit (Eq.~\ref{eq:circumbinary_stability_HW}) prevents planets from orbiting with  $a_{\rm p} \lesssim 2.5 a_{\rm bin}$, and hence $T_{\rm p}\lesssim 4 T_{\rm bin}$. Second, there is an apparent paucity of circumbinary planets orbiting the tightest eclipsing binaries ($T_{\rm bin} < 7$ days). This is shown in Fig.~\ref{fig:circumbinary_EB_histogram}, where the histogram of the {\it Kepler} eclipsing binaries has a median of 2.8 days. If planets were distributed irrespective of binary period, at least twice as many should have been discovered \citep{martin14,armstrong14}. Such tight binaries are not believed to form in situ,  but rather at wider separations followed by a process of high-eccentricity Kozai-Lidov under the influence of a misaligned third star, followed by tidal friction \citep{harrington68,mazeh79,eggleton01,tokovinin06,fabrycky07,naoz14,moe18}. This formation pathway for very tight binaries has been used to explain the dearth of observed planets around them \citep{munoz15,martinetal15,hamers16,xu16}. Most planets sandwiched in this evolving, misaligned triple system  either fail to form, or become unstable during the shrinking process, or actually inhibit the binary shrinkage. Furthermore, the rare remaining planets are expected to have small mass and orbits that are long-period and misaligned, and hence harder to discover.

\begin{figure*}  
\begin{center}  
\includegraphics[width=\textwidth]{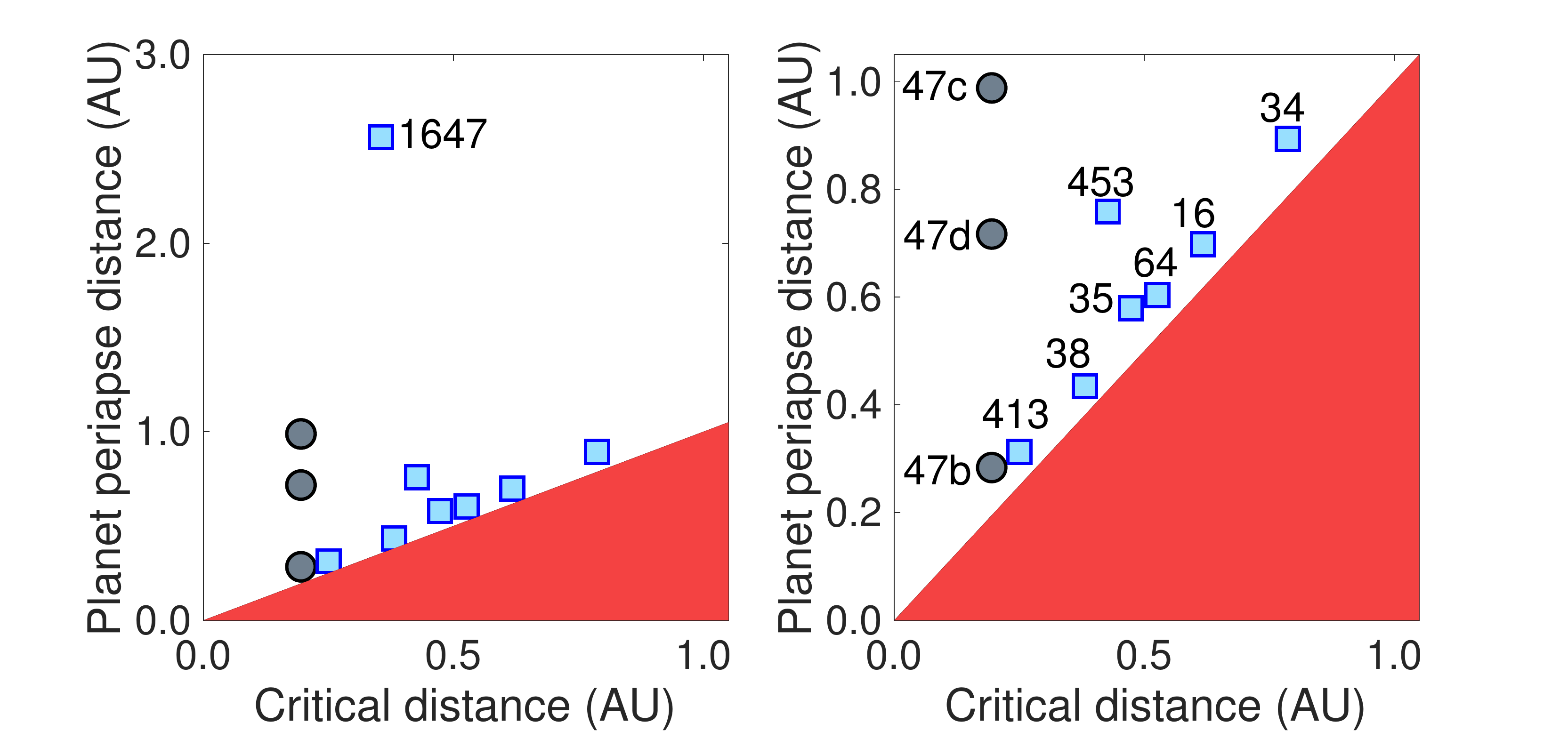}  
\caption{Circumbinary periapse distance, $a_{\rm p}(1-e_{\rm p})$, as a function of the critical distance ($a_{\rm crit}$ in Eq.~\ref{eq:circumbinary_stability_HW}) for the {\it Kepler} transiting circumbinary planets. Planetary orbits in the red zone are unstable. The multi-planet Kepler-47 system is drawn with grey circles, and the blue squares correspond to the other, single planet systems. The right plot is zoomed near the stability limit, which excludes Kepler-1647. Kepler numbers are labelled.}
\label{fig:circumbinary_stability_limit}
\end{center}  
\end{figure*}

With respect to the stability limit imposed by the dynamical influence of the binary, the circumbinary planets have generally been found as close as possible. This is demonstrated in Fig.~\ref{fig:circumbinary_stability_limit}. The planet periapse distance is plotted as an ad hoc means of including the planet eccentricity, which Eq.~\ref{eq:circumbinary_stability_HW} does not account for, although most of the known circumbinary planets have small eccentricities $e_{\rm p}<0.1$. See \citet{mardling01} for further details on the effect of the outer eccentricity. Kepler-47 is the only multi-planet system, with the innermost planet right next to the stability limit, following the trend. It would be impossible for the outer two planets in Kepler-47 to also be close to the stability limit. 

\citet{welsh14} attributed this observed pile-up to either a true preference for circumbinary planets to exist as close as possible to the stability limit, or an observing bias.   \citet{martin14} simulated the {\it Kepler} circumbinary population and could not reproduce the observed pile-up of planets with observing biases alone. The most recent Bayesian analysis of \citet{li16}, including the recently-discovered Kepler-1647 (top blue square in Fig.~\ref{fig:circumbinary_stability_limit} left), showed that there was evidence for a pile-up if this very long-period planet was an outlier of the planet period distribution. If it was instead drawn from the same distribution as all of the others then the statistical significance of the pile-up was reduced.  We note that the single planet discovered by microlensing (OGLE-2007-BLG-349L, \citealt{bennett16}) has $a_{\rm p}/a_{\rm bin}\sim 40$, far from the stability limit. The borderline RV discovery of HD 202206 \citep{correia05,fritz17} however has $a_{\rm p}/a_{\rm bin}\sim 2.3$, near the stability limit and the 5:1 resonance. Overall, more discoveries are needed, using different observing techniques with different biases.

It is seemingly difficult to form circumbinary planets in situ so close to the binary, owing to a hostile disc environment \citep{paardekooper12,lines14}. The favoured theory is an inwards migration of the planets followed by a parking near the stability limit \citep{pierens13,kley14}, where the disc is expected to have been truncated \citep{artymowicz94}.

%


\subsubsection{The occurrence rate of circumbinary planets}\label{subsubsec:abundance}

\begin{figure}  
\captionsetup[subfigure]{labelformat=empty}
\begin{center}  
	\begin{subfigure}[b]{0.49\textwidth}
		\includegraphics[width=\textwidth]{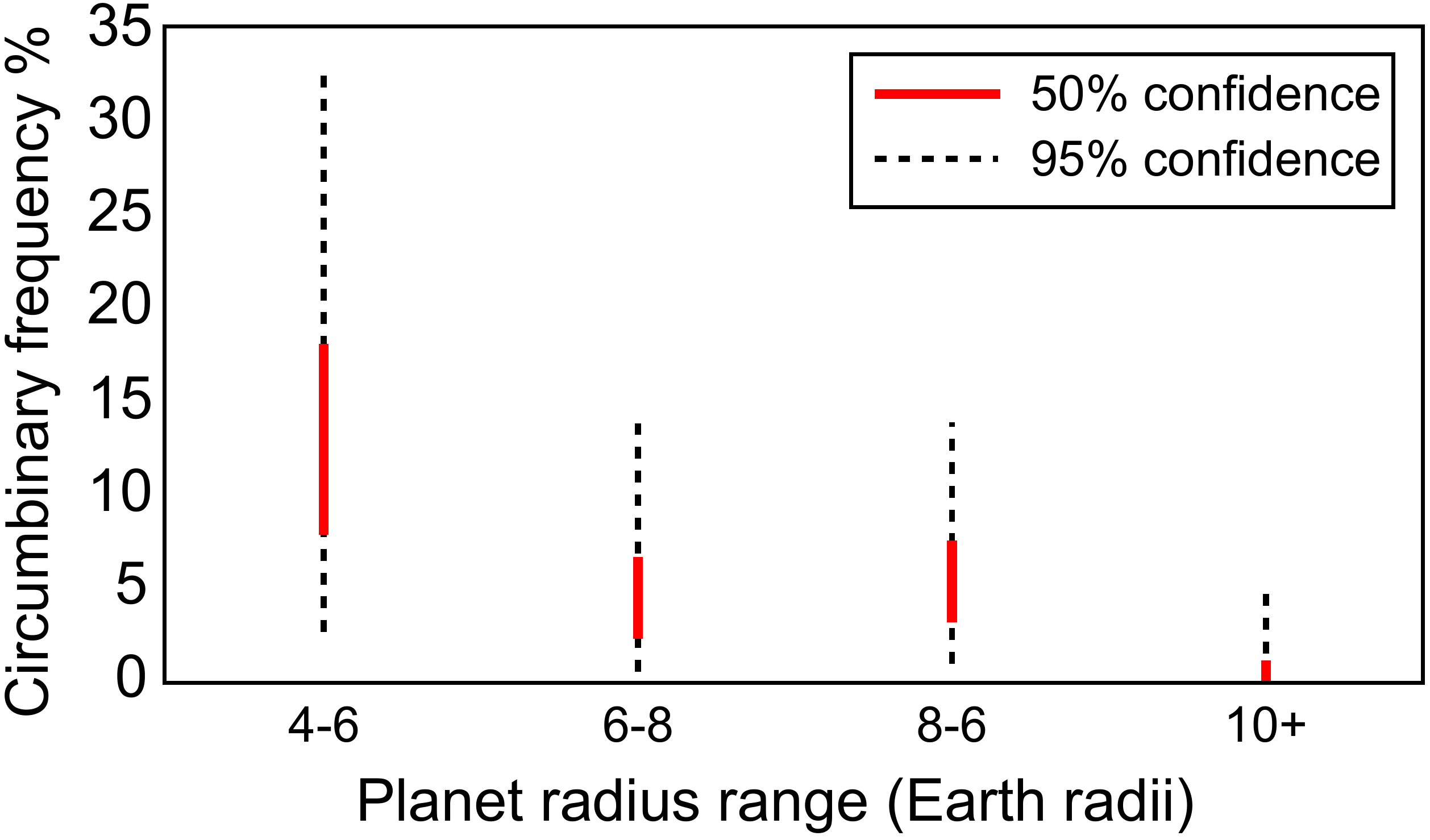} 
		\label{fig:armstrong_abundance_size} 
	\end{subfigure}	
	\begin{subfigure}[b]{0.49\textwidth}
		\includegraphics[width=\textwidth]{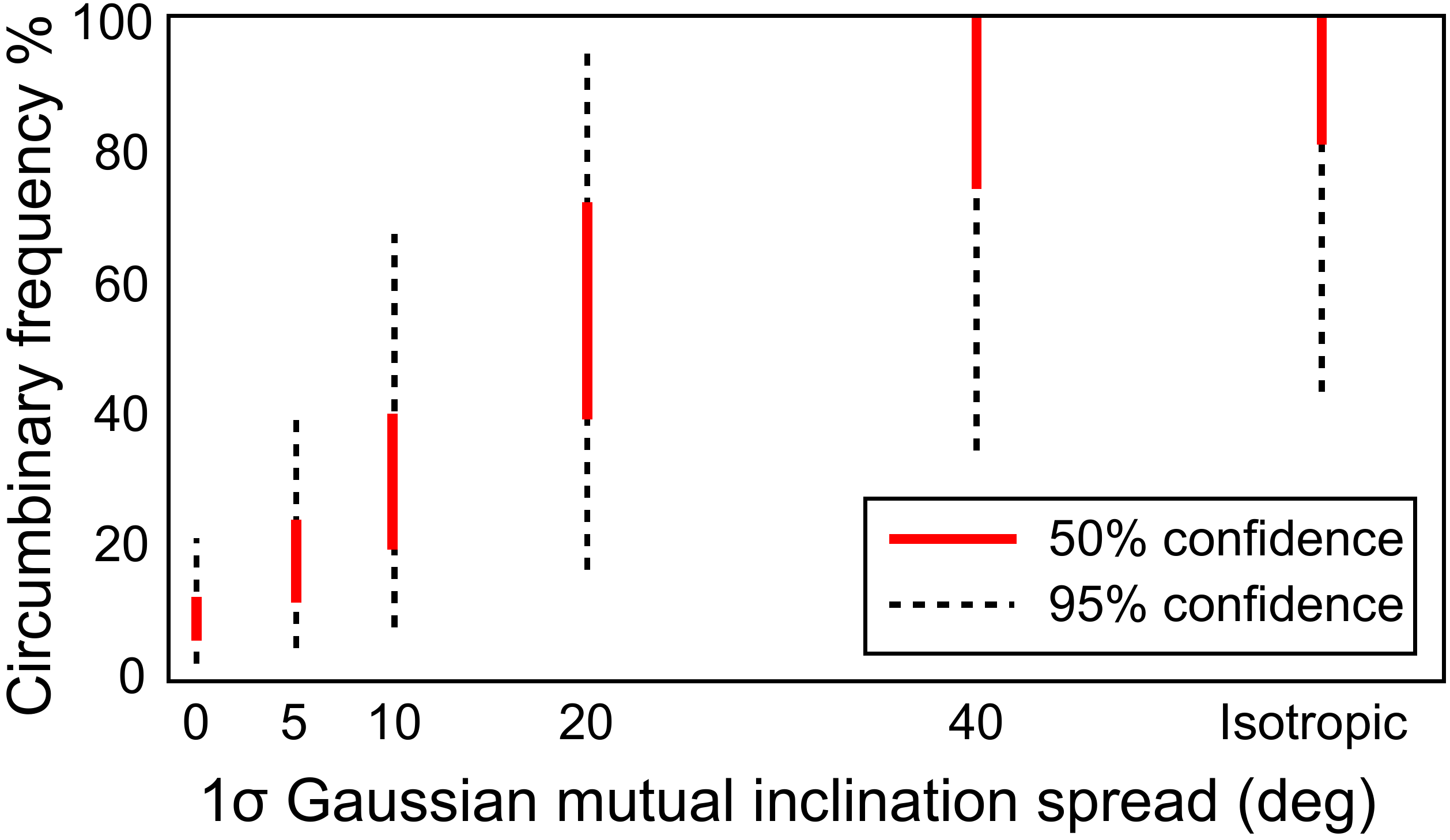}  
		\label{fig:armstrong_abundance_incl}
	\end{subfigure}
	\caption{Occurrence rate of circumbinary planets orbiting within 10.2 times the binary orbital period. Left: planets drawn from a Gaussian distribution of mutual inclinations with a $5^{\circ}$ standard deviation, as a function of the planet radius. Right: planets between 4 and 10 $R_{\oplus}$ as a function of the  standard deviation of the Gaussian mutual inclination distribution. In all cases the Gaussian distribution is convolved with an isotropic distribution (i.e. uniform in $\cos I_{\rm p}$). See \citet{armstrong14}. for more details.}
	\label{fig:armstrong_abundance}  
\end{center}  
\end{figure} 



The first estimate for the circumbinary occurrence rate was made in the discovery paper of Kepler-34 and -35 by \citet{welsh12}. They used a simple geometric approach with static orbits to calculate that for the one Kepler-16, -34 and -35 that was observed transiting an eclipsing binary, another 5, 9 and 7 similar planets should exist that did not transit. Based on 750 eclipsing binaries being analysed, they estimated the circumbinary frequency as  $(5+9+7)/750 = 2.8\%$. This was expected to be an underestimate given that the search was not exhaustive at that point.

 The studies of \citet{martin14} and \citet{armstrong14} calculated the frequency of circumbinary planets as a function of the underlying distribution of the alignment between binary and planetary orbits. All of the systems discovered so far are flat to within $\sim 4^{\circ}$. This is similar to the Solar System and multi-exoplanet systems around single stars \citep{fabrycky14}. However, the the detection efficiency of misaligned circumbinary planets is reduced; whilst they may still pass the binary orbit, they will often miss transits, creating a sparse transit signature which is hard to identify.

\citet{martin14,armstrong14} noted that any abundance deduced based on the coplanar sample would therefore only be a minimum abundance, as a highly misaligned sample of planets could not be ruled out. \citet{martin14} simulated the {\it Kepler} detection yield for a suite of hypothetical circumbinary distributions, which was then compared with the actual {\it Kepler} findings. The tested distribution which best matched the {\it Kepler} discoveries had a 10\% minimum frequency of gas giants. The more comprehensive study by \citet{armstrong14} used an automated algorithm was made to search the {\it Kepler} eclipsing binary light curves for transit signals of circumbinary planets. Its sensitivity was limited to gas giants ($\gtrsim4R_{\oplus}$). The algorithm was tested on all detached {\it Kepler} eclipsing binary light curves, searching for both real planets and injected fake transit signals. By quantifying the detectability of planets in each eclipsing binary light curve, \citet{armstrong14} derived a minimum occurrence rate that matched the $\sim 10\%$  calculation by \citet{martin14}. Both studies are higher than the initial  $\sim 3\%$  calculation by \citet{welsh12}, but the present sample size is too small to rule out this lower value.

In Fig.~\ref{fig:armstrong_abundance} (left) the \citet{armstrong14} occurrence rate the  frequency is broken down into different radius intervals. There is a decreased  frequency for larger planets, in line with what is known for single stars. Note that Kepler-1647 had not been confirmed at the time  of their analysis, and is 11.9 Earth radii. Figure~\ref{fig:armstrong_abundance} (right) demonstrates how the true frequency of circumbinary planets is a function of the underlying distributions of the alignment between the binary and planet orbital planes.

A giant circumbinary planet frequency of 10\% would be  compatible with what is seen around single stars at similar periods \citep{howard10,mayor11,petigura13}. This hints that the formation of gas giants might be similar around one and two stars.  Furthermore, the existence of a highly misaligned population of circumbinary planets would be indicative of an even higher abundance when compared with single stars, posing curious questions to planet formation theories. 

Most recently, \citet{li16} suggested that the existing discoveries can actually be used to deduce a true mutual inclination distribution of just a few degrees. However transit discovery methods that are sensitive to  highly misaligned planets  (e.g. $\gtrsim20^{\circ}$) are yet to be demonstrated.  Overall, more circumbinary discoveries are required to draw any firm conclusions.

 \citet{klagyivik17} searched for circumbinary planets using data from the {\it CoRoT}  mission, which preceded {\it Kepler}. The shorter {\it CoRoT} observing timespans between 30 and 180 days limited the search sensitivity to $P_{\rm p}< 50$ days and $P_{\rm bin} < 10$ days. No discoveries were made, but within this period range the Jupiter- and Saturn-sized circumbinary  frequency was constrained to $<0.25\%$ and $0.56\%$, respectively. This is much smaller than seen for comparable planets around single stars, but fitting with the dearth of circumbinary planets around tight binaries found in the {\it Kepler} mission.

Efforts have also been made to quantify the circumbinary  frequency at wider separations. The SPOTS survey conducts direct imaging on  young spectroscopic binaries to search for outer companions \citep{thalmann14}. The initial sample of 26 binaries has a wide spread of periods ranging from 1 day up to 40 years. The latest work in \citet{bonavita16} has been to combine observations taken in SPOTS with those already existing in the literature. No confirmed detections were made, but the frequency of planets between 2 and 15 $M_{\rm Jup}$ between 10 and 1000 AU was confined to $<9$\% with 95\% confidence. For comparison, \citet{bowler16} analysed single stars and made  a much more precise occurrence rate calculation of  $0.8^{+1.0}_{-0.6}\%$ for $5-13M_{\rm Jup}$ planets in wide $10-1000$ AU orbits. Surveys of massive, long-period circumbinary planets are therefore comparatively in their infancy.

The imaging surveys have focused on young systems and the {\it Kepler} results have been for main sequence binaries. Contrastingly, the method of ETVs has typically focused on evolved, post-common envelope binaries with $P_{\rm bin}<1$ day. \citet{zorotovic13} find that roughly 90\% of such binaries have observed ETVs, which could be interpreted as planets. This is roughly 10 times larger than seen in {\it Kepler} or the SPOTS survey. This indicates that ETVs observed are unlikely to all be of planetary origin, and likely include false positives such as the Applegate mechanism \citep{applegate92}. Alternatively, there would need to be a highly effective means of second generation planet formation after the evolution of the inner binary \citep{perets10,bear14}.

\subsection{Circumprimary and circumsecondary planets}\label{subsec:circumprimary}

Methodologically, there are two approaches to finding circumstellar planets in binaries. First, a binary may already be known and then a search is made for interior planets, for example the \citet{eggenberger06,toyota09} surveys. Alternately, a planet may already been known and then there is a search for outer stellar companions. The latter approach is favoured in the literature, because finding an additional star is simply an easier task than finding an additional planet.

In Fig.~\ref{fig:all_planets} we see that most of the circumstellar planets in binaries have been discovered by transits and RVs. The binaries themselves are generally discovered with RVs, imaging and astrometry, sometimes in combination. In this figure only part of the stable parameter space is well-populated. There is a lack of wide-orbit planets ($a_{\rm p}\gtrsim10$ AU). This can be explained by the difficulty in finding planets so far from their host star, particularly with the RV and transit techniques. This may change in the near future as direct imaging continues to improve.

There is also a reduced number of planets around binaries with $a_{\rm bin}<50$ AU (mean of the log-normal binary separation distribution, \citealt{raghavan10}). We know of 17 circumstellar planets in tighter binaries, compared to 101 planets in wider systems. The tightest binary known to host a circumstellar planet is 5.3 AU (KOI-1257, \citealt{santerne14}), although continued RV follow-up is on going to better characterise the outer orbit. There also exist some borderline binary cases with brown dwarf secondary ``stars'' in even tighter orbits (WASP-53 and WASP-81, \citealt{triaud17},  but not included in Fig.~\ref{fig:all_planets}). Tight binaries may not be resolvable by imaging surveys, but they are the easiest to find by the RV technique. Additionally, {\it Kepler} survey has provided almost 3,000 eclipsing binaries, with periods ranging from less than a day to several hundred (Fig.~\ref{fig:circumbinary_EB_histogram}), but none are known to host circumstellar planets.

 We first review the multiplicity of planet-hosting stars, particularly as a function of binary separation, for example like the aforementioned dearth of planets in $a_{\rm bin}<50$ AU binaries. Comparisons are also made with the multiplicity of stars in general.  The special class of hot Jupiters is then treated separately, before finally summarising some of the difficulties and caveats in the studies of circumstellar planets in binaries.

\subsubsection{The stellar multiplicity of planet hosts }\label{subsubsec:multiplicity}

\begin{figure*}  
\begin{center}  
\includegraphics[width=0.8\textwidth]{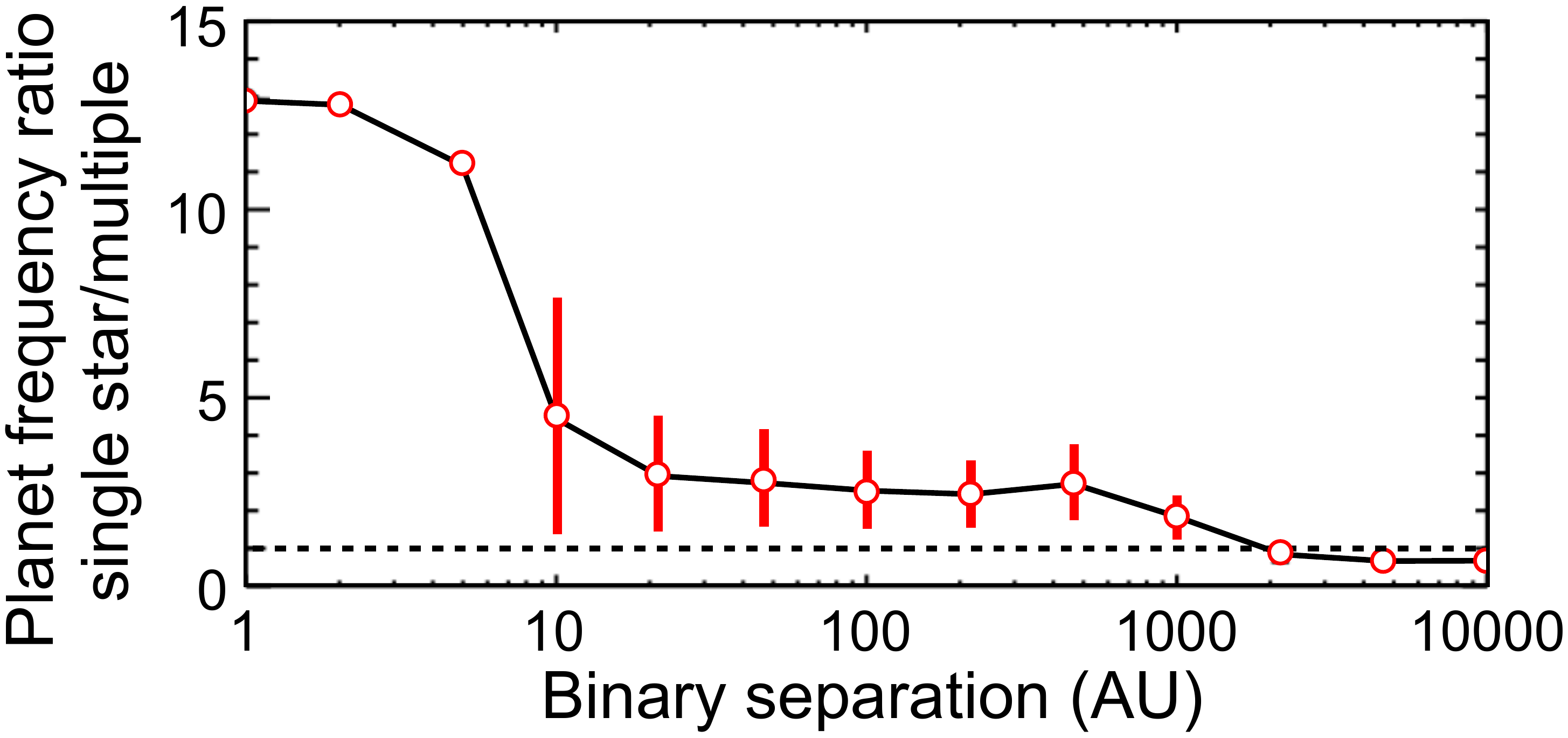}  
\caption{Ratio of the planet  frequency in single star systems to that in multi-star systems as a function of the separation to the stellar companion, taken from \citet{wang2}  based on imaging of KOIs. Error bars come  from Poisson statistics, but are not calculated for the first three data points due to a lack of detected stellar companions. Error bars for the last three data points are invisibly small. The dashed line is a ratio of 1.}
\label{fig:wang2}
\end{center}  
\end{figure*}

 There have been two main sources of planet-hosting stars around which stellar companions were searched. Earlier studies used planets discovered by RVs. More recent work has used Kepler Objects of Interest (KOIs), i.e. transiting planet candidates. The two samples typically have vastly different planet properties, sample biases and observational sensitivities to outer companions. Consistency in the stellar multiplicity rates is therefore not necessarily expected. However, some of the same trends have been seen in both samples.

One of the first large studies was conducted by \citet{eggenberger07}. A sample was constructed of 130 RV target stars, half of which were known to host a gas giant planet and the other half used as a control sample. Direct imaging was used to uncover outer stars.  The control sample multiplicity was 18\%, almost double that of the planet-host sample which had a multiplciity rate of 10\%. \citet{eggenberger11} showed that whilst the planet hosts have a lower rate of stellar companions than field stars within 100 AU, there was no discernible difference for companions between 100 and 200 AU. The independent \citet{desidera07} imaging survey also recovers the detrimental impact of binaries tighter than 100 AU. \citet{ginski12,ginski16} surveyed 125 RV  planet hosts and calculated an overall smaller multiplicity of $5.6\%$ based on confirmed stellar companions, but this percentage raises to $9-10\%$ if unconfirmed companions were included. 

 \citet{ngo17} compared the distribution of mass, period and eccentricity of RV planets around stars within and without stellar companions within 6 arcsecs. They found no discernable difference. The complementary survey of \citet{moutou17} observed multi-stellar systems with wider separations. It was found that that eccentric RV planets are more likely to exist in a binary than circular RV planets, potentially as a consequence of dynamical perturbations (see simulations by \citealt{kaib13}).

\citet{wang1}  combined imaging and spectroscopic measurements of KOIs in the search for stellar companions. They demonstrated a paucity of planets in tight binaries ($\lesssim 20$ AU), for which the multiplicity of planet hosts was roughly three times less than for field stars.  The follow-up study of \citet{wang2} was extended to to wider binary separations. They found a small depletion of planets in binaries as wide as 1500 AU, but only at 1-2$\sigma$ significance. In Fig.~\ref{fig:wang2} we plot their calculated ratio of the planet  frequency in single star systems to that in multi-star systems. This matched the later work of \citet{kraus16} to also directly image KOIs calculated the suppression of planets in tight binaries ($< 50^{+49}_{-23}$ AU) by a factor of 3 compared to the  frequency around single stars or wider binaries.  Accounting for both the paucity and the rate of stellar multiplicity in field stars, it was deduced that one fifth of all solar-type stars are unable to host exoplanets, owing to a detrimental effect of a binary companion.

 The study of \citet{horch14} similarly targeted KOIs with direct imaging, but at a lower spatial resolution. Consequently, they were typically sensitive to wider binaries than the previously-mentioned KOI surveys. They calculated a multiplicity rate of (37\% $\pm$ 7\%) and (47\% $\pm$ 19\%), based on the work done using the WIYN 3.5 m and Gemini North 8.1 m telescopes, respectively. These numbers are similar to the multiplicity of field stars ($\sim 50\%$, \citealt{duquennoy91,raghavan10}). The \citet{horch14} results are consistent with those from \citet{wang1,wang2,kraus16} for wide binaries, i.e. there is minimal or no impact of wide stellar companions ($\gtrsim 100$ AU) on planet occurrence.


A follow-up imaging survey of \citet{wang3} focused on solely giant planet KOIs,  and hence may be more easily compared with RV-discovered planets. They discerned that the multiplicity of planet hosts was  depleted to $0^{+5}_{-0}\%$ for binaries within 20 AU  when compared to $18\pm2\%$ for field stars.  Contrastingly, \citet{wang3} discovered a surprising increase in the multiplicity of planet hosts to $34\pm8\%$ for binaries between 20 and 200 AU, which is significantly higher than  the field star multiplicity of $12\pm2\%$.  This is a result not seen in studies of RV planet hosts and warrants further investigation, particularly given the potential consequences on planet formation. For binaries wider than $\sim 200$ AU the multiplicity rate of field stars and planet hosts was comparable, as found by other authors.

\begin{figure*}  
\begin{center}  
\includegraphics[width=0.7\textwidth]{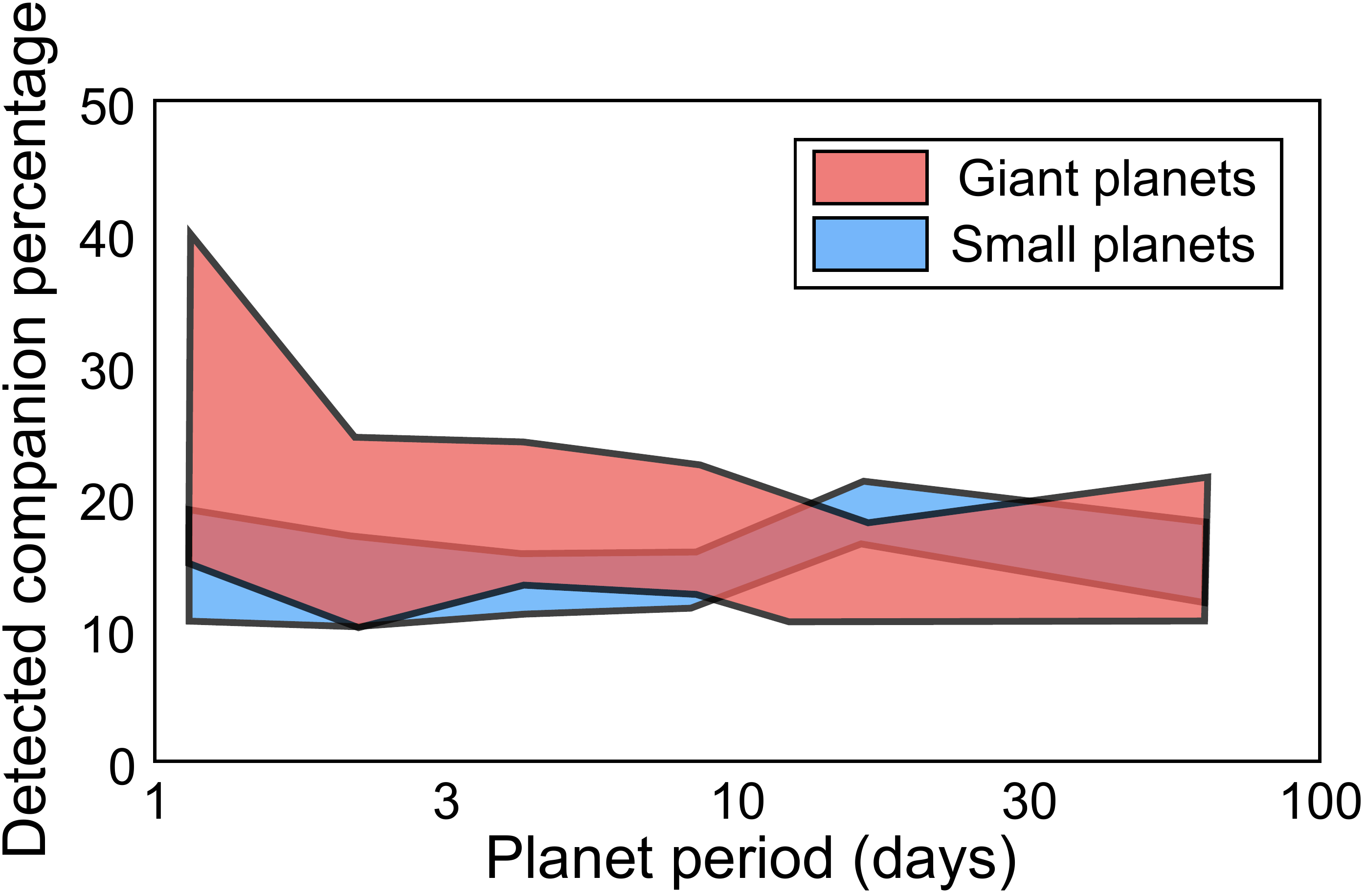}  
\caption{Binary percentage of giant planets ($>3.9R_{\oplus}$) and small planets based on imaging of {\it Kepler} candidates by \citet{ziegler17a}. Error bars are $1\sigma$.}
\label{fig:ziegler_2017}
\end{center}  
\end{figure*}

 Since 2012 the Robo-AO survey has conducted adaptive optics follow-up of hosts of {\it Kepler} planet candidates, with a sensitivity out to $4''$. This work has been published in a series of four papers \citep{law14,baranec16,ziegler17a,ziegler17b}. In Fig.~\ref{fig:ziegler_2017} we show their comparative stellar binary rates for hosts of giant ($>3.9R_{\oplus}$) and smaller planets. At short periods less than $\sim 10$ days there is a marginal increase in stellar multiplicity for giant planets. No statistically significant differences are seen at longer planet periods.

The presence of a close binary companion has strong implications for planet formation theories. It is predicted that the protoplanetary disc will be truncated \citep{artymowicz94} and that its conditions will be less favourable for planet formation by both gravitational collapse and core accretion \citep{nelson00,mayer05}. There may also be an ejection of formed planets \citep{zuckerman14}. Observations of protoplanetary discs also show evidence for decreased lifetimes in $<100$ AU binaries \citep{kraus12,daemgen13,daemgen15,cheetham15}.

\subsubsection{Hot Jupiters in stellar binaries}\label{subsubsec:hot_jupiters}

The existence and properties of ``hot Jupiters'' - giant planets on orbits of just a few days - have confounded us ever since the first discovery of 51 Peg \citep{mayor95} (see chapter by Santerne). The environment at such close proximity of the stars has classically thought to be a hinderance to planet formation (\citealt{pollack96,rafikov06}, but see also \citealt{boley16,batygin16}). Alternatively, the giant planet forms farther out in the disc before migrating inwards. Several different migration mechanisms have been proposed, such as disc migration \citep{goldreich79,lin79,ward97,masset03}, planet-planet scattering \citep{weidenschilling96,rasio96,chatterjee08,beauge12} and {Kozai-Lidov cycles plus tidal friction} \citep{innanen97,wu03,fabrycky07,naoz12}. 

It has been observed that $\sim 30\%$ of hot Jupiters exist on orbits that are misaligned or even retrograde with respect to the spin of the host star (\citealt{hebrard08,winn09,triaud10} and the chapter by Triaud). This may be a fingerprint of Kozai-Lidov cycles acting on the inner orbit. Alternatively, the misalignment  distribution may be a reflection of the tilting of the protoplanetary disc \citep{lai14,spalding14,spalding15,matsakos17}  or planetary engulfment \citep{matsakos15}.  A massive outer body is often implicated in these theories, and hence stellar binaries have been targeted as an explanation for hot Jupiters.

\begin{figure*}  
\begin{center}  
\includegraphics[width=0.8\textwidth]{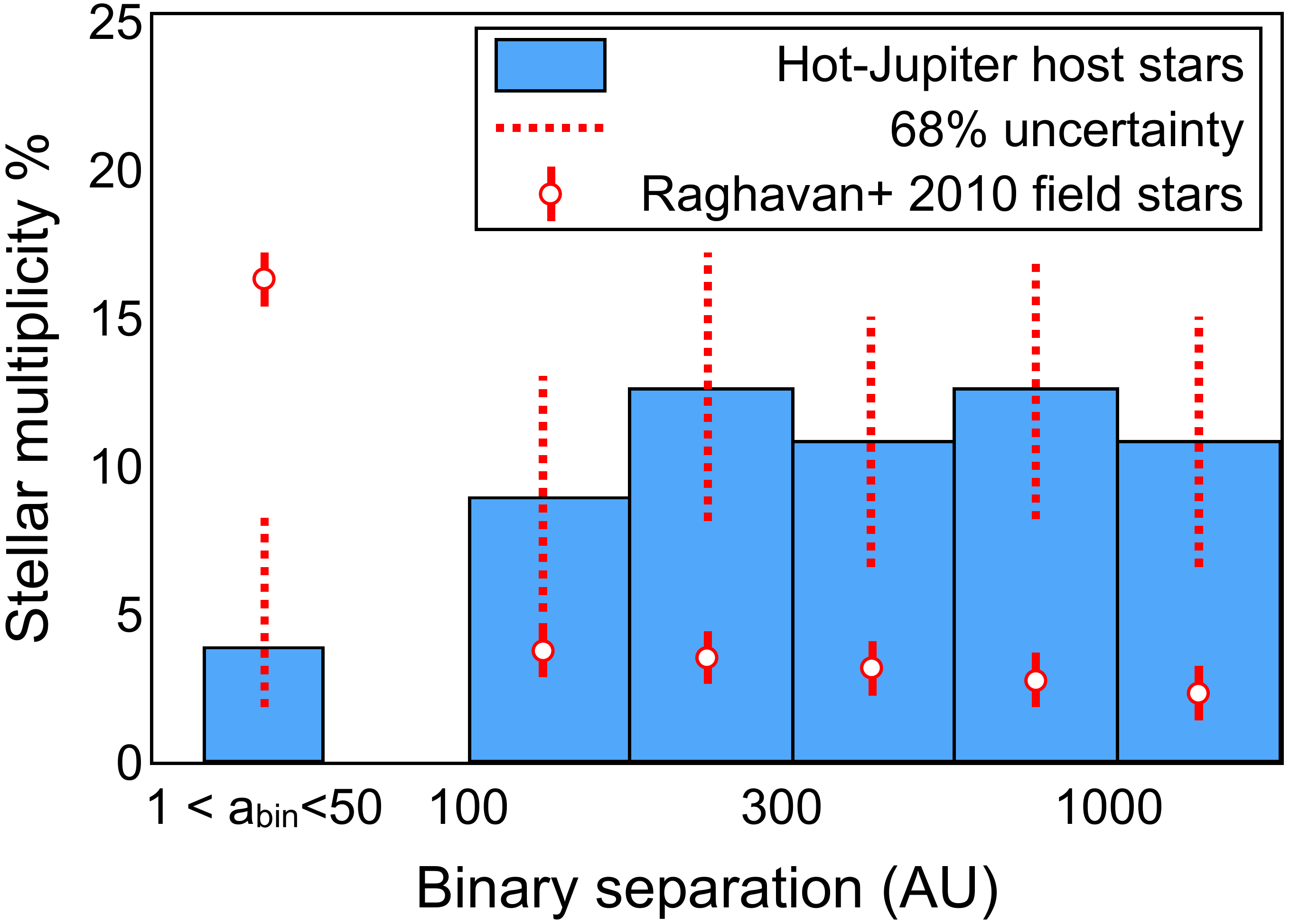}  
\caption{ Completeness-corrected stellar multiplicity rate of hot Jupiter hosts compared with field stars from \citet{ngo16}. The multiplicity is calculated in five uniform bins in log space between 100 and 1778 AU, corresponding to the sensitivity of imaging, and an additional bin between 1 and 50 AU for the closer companions detectable by RVs.}
\label{fig:friends_of_hot_jupiters}
\end{center}  
\end{figure*}

The ``friends of hot Jupiters'' survey has searched for outer companions to hot Jupiters drawn predominantly from the {\it WASP} and {\it HAT} photometric surveys. The results have been presented in a series of papers \citep{knutson14,ngo15,piskorz15,ngo16}. Radial velocities are used to search for close companions ($a_{\rm bin}<50$ AU), whereas direct imaging probes farther bodies. Two contrasting results were discovered, as shown in Fig.~\ref{fig:friends_of_hot_jupiters}. For $a_{\rm bin}$ between 1 to 50 AU the multiplicity of hot Jupiter hosts is $3.9^{+4.5}_{-2.0}\%$, which is roughly four times less than what is seen for field stars. The presence of a close stellar companion is seemingly detrimental to the existence of a hot Jupiter. On the other hand, hot Jupiters are seen to have wider stellar companions (50 - 2000 AU) at a rate of $47\pm7\%$, which is  three times larger than what is seen for field stars. Note that in Fig.~\ref{fig:friends_of_hot_jupiters} this high multiplicity is split into five separation bins. 

The \citet{evans16} imaging survey of predominantly {\it WASP} and {\it CoRoT} hot Jupiters calculated a multiplicity rate of $38^{+17}_{-13}\%$.  Since their survey was typically only sensitive to companions farther than 200 AU, this rate is expectantly slightly lower than that calculated by the \citet{ngo16} survey, whose imaging was sensitive to companions as close as 50 AU. The work of \citet{daemgen09,faedi13,adams13,bergfors13} have also reported multiplicity rates that are slightly smaller than \citet{ngo16}, but only calculated using a small sample of $\sim 15-20$ hot Jupiters. The Robo-AO survey results from \citet{ziegler17a} (Fig.~\ref{fig:ziegler_2017}) demonstrate a heightened stellar multiplicity of stars hosting hot Jupiters compared with those hosting hot small planets.

 The inherent rarity of hot Jupiters does mean that the current statistics significant have room for improvement. This problem will hopefully be overcome by {\it TESS} (see chapter by Ricker) and {\it PLATO} (see chapter by Rauer \& Heras). Improvements in imaging and results from the {\it GAIA} astrometric survey (see chapter by Sozzetti \& Bruijne) will aid the detection of stellar companions.

On the surface, a heightened stellar multiplicity of hot Jupiter hosts gives credence to the idea of Kozai-Lidov migration. However, two problems have been uncovered. First, a stellar companion is not always sufficient to induce Kozai-Lidov cycles, as a faster secular effect will quench them, for example apsidal precession induced by tides and general relativity. The timescale of Kozai-Lidov cycles  increases for farther companions, like the ones generally found around hot Jupiters. \citet{ngo16} calculated that only $\sim20\%$ of their  surveyed hot Jupiters could have formed by Kozai-Lidov migration. This result is compatible with earlier simulations of \citet{wu07,naoz12,petrovich15}. A second problem with the Kozai-Lidov migration scenario is that \citet{ngo15} found that the stellar multiplicity rate was not correlated with the misalignment of hot Jupiters.

 It is possible, however, that Kozai-Lidov migration is typically caused not by stellar companions to hot Jupiters but rather planetary companions. This has been investigated theoretically \citep{naoz11} and observationally \citep{bryan16}, but is beyond the scope of this review.

\subsubsection{Caveats and difficulties}\label{subsubsec:difficulties}

 Overall, it is hard to quantitatively compare the derived stellar multiplicity rates between different surveys. Difficulties arise from inconsistent sample selection, e.g. planets detected by RVs or transits. Radial velocity exoplanet surveys have historically avoided binary systems, owing to the threat of spectral contamination. This is expected to be the main reason why multiplicity rates of RV-discovered planets are lower than that for transiting planets. Planets found using RVs are also typically larger and on longer periods, and both properties are seemingly connected to the influence of stellar companions.

Even for a single detection method such as transits, a difference in precision and observational timespan (e.g. {\it Kepler} verses {\it WASP}/{\it HAT}) biases the size and period distribution of the planets and hence potentially the deduced rate of stellar multiples. An additional effect, which has not been explored here, is the effect of the host star mass. This has been known to affect both the stellar multiplicity rate and semi-major axis distribution of binaries \citep{raghavan10}, as well as the planet occurrence rate around single stars \citep{johnson10}. 

Another difficulty is the presence of the Malmquist bias, which is a preferential selection of brighter targets within astronomical surveys. Since multi-star systems have more flux contributions than single stars, they may be overrepresented in some surveys, skewing statistics. See \citet{kraus16} for further discussion on the Malmquist bias in multiplicity studies, and also \citet{wang2,wang_other,ginski16} for a more in depth discussion on other challenges.

\section{Summary of observed trends}\label{sec:summary}

 Listed here are the most compelling observational trends so far.

\vspace{0.4cm}

Circumbinary (p-type) planets:

\begin{itemize}
\item Giant circumbinary planets around moderately wide binaries ($P_{\rm bin}\sim 7-41$ days) are found at a similar frequency to similar sized planets around single stars, hinting at a similar formation efficiency around one and two stars.
\item There is a dearth of circumbinary planets around tighter binaries, which is seen as evidence for the existing theory of tight binary formation formation via Kozai-Lidov cycles under the influence of a third star, plus tidal friction.
\item There is an over-abundance of circumbinary planets near the orbital stability limit. This may be indicative of inwards migration within the protoplanetary disc before a parking mechanism stops the planets near the inner hole in the disc which has been carved out by the binary.
\end{itemize}

Circumstellar (s-type) planets in binaries:

\begin{itemize}
\item When marginalized over all planet sizes, tight stellar companions ($\lesssim 50$ AU) are $\sim 3$ times less likely to be found around exoplanet hosts than field stars. This suggests a ruinous influence of a tight binary on planet formation and/or survival.
\item For wider binaries the planet host and field star multiplicity rates are similar, so additional stars at these separations are seemingly too distant to influence the planets. This is again for planets of all sizes.
\item Hot Jupiters have a $\sim 3$ times heightened stellar multiplicity rate compared to field stars, but only for wide ($ \gtrsim 50$ AU) binaries. This may be indicative of a nurturing influence of a wide stellar companion on hot Jupiters.
\end{itemize}

\section{Cross-References}

\begin{itemize}
\item{Two Suns in the Sky: The Kepler Circumbinary Planets $-$ Welsh, W. \& Orosz, J.}
\item{Circumbinary Planets Around Evolved Stars $-$ Marsh, T.}
\item{The Way to Circumbinary Planets $-$ Doyle, L. \& Deeg, H.}
\item{The Rossiter/McLaughlin Effect in Exoplanet Research $-$ Triaud, A.}
\item{Hot Jupiter Populations from Transit and RV Surveys $-$ Santerne, A.}
\item{Space Astrometry Missions for Exoplanet Science: Gaia and the Legacy of Hipparcos $-$ Sozzetti, A. \& Bruijne, J.}
\item{Space Missions for Exoplanet Science: TESS $-$ Ricker, G.}
\item{Space Missions for Exoplanet Science: PLATO $-$ Rauer, H. \& Heras, A.}
\end{itemize}

\begin{acknowledgement}
Thank you to Dave Armstrong, Sebastian Daemgen, Dan Fabrycky, Elliott Horch, Adam Kraus, Henry Ngo, Richard Schwarz and Amaury Triaud for expert insights on earlier versions of the manuscript. I also thank section editor Natalie Batalha for her thorough review, and book editors Hans Deeg and Juan Antonio Belmonte for giving me the opportunity to write this chapter. Finally, I acknowledge funding and support from the Swiss National Science Foundation, The University of Chicago and the Unversit{\'e} de Gen{\`e}ve. \end{acknowledgement}



\end{document}